\documentclass[%
 reprint,
 amsmath,amssymb,
 aps,
]{revtex4-2}

\usepackage{physics}
\usepackage{graphicx}
\usepackage{dcolumn}
\usepackage{bm}

\begin{document}

\preprint{APS/123-QED}

\title{Pulse-level Scheduling of Quantum Circuits for Neutral-Atom Devices}

\author{Richard Bing-Shiun Tsai}
\altaffiliation[]{These authors contributed equally to this work.}

\author{Henrique Silvério}
\altaffiliation[]{These authors contributed equally to this work.}

\author{Loïc Henriet}%
\email{loic@pasqal.com}

\affiliation{%
 Pasqal, 7 rue Léonard de Vinci, 91300 Massy, France
}%

\date{\today}

\begin{abstract}
We show how a pulse-level implementation of the multi-qubit gates in neutral-atom device architectures allows for the simultaneous execution of single- and multi-qubit gates acting on overlapping sets of qubits, in a mechanism we name absorption. With absorption as a foundation, we present an algorithm to schedule the execution of a quantum circuit as a pulse sequence on a neutral-atom device with a single channel for single- and multi-qubit gate execution. For any quantum circuit of practical relevance, we observe that the algorithm results in an optimal utilization of the available resources that cannot be surpassed by a different scheduling strategy. Our benchmarks against a custom scheduler attempting to maximize parallelization at the gate level show the time gained by the pulse-level scheduler is proportional to the depth and is most pronounced for quantum circuits with fewer qubits.
\end{abstract}

\maketitle


\section{\label{sec:intro}Introduction}

In the past decade, Quantum Processing Units (QPUs) made of individual atoms trapped in arrays of dipole traps have been used extensively to address quantum simulation problems\,\cite{Browaeys2020}. Those successes were achieved by leveraging the high programability of the quantum dynamics through tunable Hamiltonians. This notably includes the exploration of the formation of antiferromagnetic order in the Ising model\,\cite{Song21,Scholl2021,Ebadi2021}, or the probing of exotic states of matter\,\cite{Leseleuc2019,Semeghini2021,Bluvstein21}.\\
In addition to the analog mode where the time evolution of qubits is described by the direct manipulation of the Hamiltonian of the system, control over the device can be realized at the digital level, where
the dynamics is encoded in quantum circuits, i.e. sequences of discrete quantum gates. Already more than 20 years ago, it was proposed that the strong interactions between
Rydberg atoms could be harnessed to implement fast and robust entangling gates between neutral atoms\,\cite{Fast_QGates,Lukin2001}. The experimentally achieved fidelities have progressively increased over the past few years\,\cite{Isenhower2010,Jau2016,Lukin-Alt-CZ,Madjarov2020}, and reached now a level that allows for the implementation of complete quantum algorithms comprising numerous consecutive gates\,\cite{Graham2022,Bluvstein2022}. These rapid advances motivate the need for optimizing the way quantum circuits are implemented on hardware.\\

In this paper, we propose an efficient pulse-level scheduling of quantum circuits on neutral-atom hardware. We introduce the concept of gate absorption, which corresponds to the simultaneous execution of a pulse associated to a single-qubit gate and one of the pulses of a multi-qubit gate involving the same qubit. This mechanism enables the design of shorter pulse sequences, which would lead to improved performances for algorithm execution. The manuscript is structured as follows: in Section\,\ref{sec:naqc}, we briefly recall details about the use of neutral atom QPUs as digital quantum computers. The main ideas behind our efficient pulse-scheduling method are presented in Section\,\ref{sec:qc-to-ps}, in which we describe how to efficiently write a quantum circuit in terms of pulse sequences. The general algorithm describing the procedure is presented in Section\,\ref{sec:algo} and numerical benchmarks are shown in Section\,\ref{sec:bench}, comparing the benefits of our approach with respect to a gate-level scheduling algorithm.

\section{\label{sec:naqc}Neutral-atom devices as Digital Quantum Computers}


\subsection{Qubits and Quantum Gates}

In neutral-atom quantum devices, qubits $\ket{0}$ and $\ket{1}$ are typically encoded in two hyperfine atomic ground states (F = 1 and F = 2). Coherent transitions between these two energy levels can be driven through Raman transitions\,\cite{Fast_QGates}. Arbitrary single-qubit gates can be achieved by combining a single resonant pulse of tunable amplitude and/or duration with a change in the phase reference frame of the qubit \cite{Silverio2022pulseropensource, IBM_VZ-gates}.

Hyperfine states have very long lifetimes and barely interact with their environment, making them robust against decoherence. In order to generate entanglement in the system, a third state is required: the \textit{Rydberg state}, $\ket{r}$. The coherent coupling between ground and Rydberg states can be achieved through optical driving with a laser system. When in the Rydberg state, an atom interacts with other atoms in the Rydberg state over large distances through strong Van der Waals interactions\,\cite{Gaetan2009,Fast_QGates}. For two atoms separated by a few micrometers, this interaction can be made strong enough to shift significantly the energy of the doubly excited state, preventing the excitation of two atoms at the same time under coherent driving -- an effect called Rydberg blockade. For the purposes of this paper, we consider that two atoms can interact when they are within a distance $R_b$ (the Rydberg blockade radius) of each other. 
Naturally, to execute multi-qubit gates we need a different type of channel to address the transition between the $\ket{0}$ and the $\ket{r}$ states; borrowing the nomenclature used in Ref. \cite{Silverio2022pulseropensource} we will refer to this channel as the \textit{Rydberg} channel, in opposition to the \textit{Raman} channel that is responsible for sending the pulses driving the Raman transitions between hyperfine states.

In neutral-atom devices, one example of a native two-qubit gate is the controlled-Z (CZ) gate. Although alternative protocols for its implementation exist \cite{Lukin-Alt-CZ}, here we will focus on the protocol described in Ref. \cite{Fast_QGates} leveraging the Rydberg blockade effect for two atoms addressed sequentially by a laser beam. The gate takes advantage of the fact that bringing an atom to a Rydberg state and back induces the accumulation of a phase of $\pi$.
It starts with a $\pi$-pulse on one atom, which will bring it to $\ket{r}$ if its state is $\ket{0}$ and will have no effect otherwise. Next comes a $2\pi$-pulse on the other atom, which will bring it to $e^{i\pi}\ket{0}$ if it started out in $\ket{0}$ and the first atom is not already in $\ket{r}$ --- in this particular case (i.e. the $\ket{00}$ case), the Rydberg blockade suppresses the effect of the $2\pi$-pulse. Finally, a third $\pi$-pulse deexcites the first atom back to $\ket{0}$, had it been excited to $\ket{r}$ by the first pulse. Overall, for all states where at least one of the atoms starts in state $\ket{0}$, one (and only one) $2\pi$-rotation will be effective, leading to the pair state acquiring a $e^{i\pi}$ phase. On the other hand, when the initial state is $\ket{11}$, no pulse is effective and thus no phase is acquired. Therefore, this protocol has an identical effect to that of the CZ gate (where only the $\ket{11}$ state changes to $-\ket{11}$), up to a global phase of $\pi$.

In fact, although a CZ (together with arbitrary single-qubit gates) is enough to form a universal gate set, the gate scheme is naturally extendable to multi-qubit gates on more than two qubits. In the case of a set of $N$ atoms, where each atom is within the interaction range of all the others, we can execute a multi-controlled-Z gate (MCZ) by applying a sequence of $N-1$ $\pi$-pulses on $N-1$ atoms, then the $2\pi$-pulse on the remaining atom, followed by another sequence of $N-1$ $\pi$-pulses in the inverse order they were first applied in --- meaning that the atom receiving the first $\pi$-pulse will also receive the last, the atom receiving the second $\pi$-pulse also receives the second to last $\pi$-pulse, and so on. With this scheme, the first atom that is excited to $\ket{r}$ will blockade all others from also being excited, thus making it so that all computational states will acquire a phase of $\pi$ except for the $\ket{11...1}$ state, as is the case for the CZ. 

Furthermore, it is worth noting that the effect of an MCZ, in spite of its name, does not depend on which qubit is chosen to be the target. This implies that there is freedom in choosing the order with which the first sequence of $\pi$-pulses is applied --- which consequently defines by exclusion the qubit receiving the $2\pi$-pulse. 

Throughout this paper, we shall abuse the term \textit{MCZ} to refer to a general controlled-Z gate with any number of controls, including a single one (i.e. the CZ).



\subsection{Physical Implementation}

As has been shown, quantum gates are executed by specific channels, through one or more pulses. Physically, these channels correspond to electromagnetic sources sending signals with precisely tuned frequencies to drive the desired transitions. These signals can be focused such that they reach individual qubits, rather than the ensemble of atoms.
Experimentally, it is impractical to have one of such channels, of each type, targeting an individual qubit exclusively. Instead, what is done in practice is that a small set of channels --- comprising at least one channel of each type --- is responsible for targeting multiple qubits in the system, making them effectively change their target throughout the pulse sequence. This change of target is generally not instantaneous, and is thus necessary to consider this retargeting time when converting a series of quantum gates into a sequence of pulses. Furthermore, having less channels than qubits reduces the degree at which gates can be executed in parallel, as the same channel generally cannot be executing different gates on different qubits at the same time. 

\section{\label{sec:qc-to-ps} From a Quantum Circuit to a Pulse Sequence}

To execute a quantum circuit, there comes a point where it needs to be translated into a sequence of pulses that the quantum device can perform. Before this step, it is assumed that the quantum circuit has been properly optimized and transpiled\,\footnote{We use the term \textit{transpilation} to refer to the transformations made to a quantum circuit at the gate level that ensure and leverage its execution on a specific quantum device.} at the quantum-circuit level; we assume that the circuit has been previously processed such that:
\begin{enumerate}
    \item All quantum gates in the circuit are native gates of the device.
    \item All multi-qubit gates in the circuit involve qubits capable of interacting among themselves.
    \item All consecutive single-qubit gates on the same qubit are merged into one single-qubit gate.
    \item The circuit has no redundant gates (i.e. gates whose presence never changes the outcome of the circuit).
    \item All qubits in the circuit are involved in at least one multi-qubit gate.
\end{enumerate}

\begin{figure}
\includegraphics[width=\columnwidth]{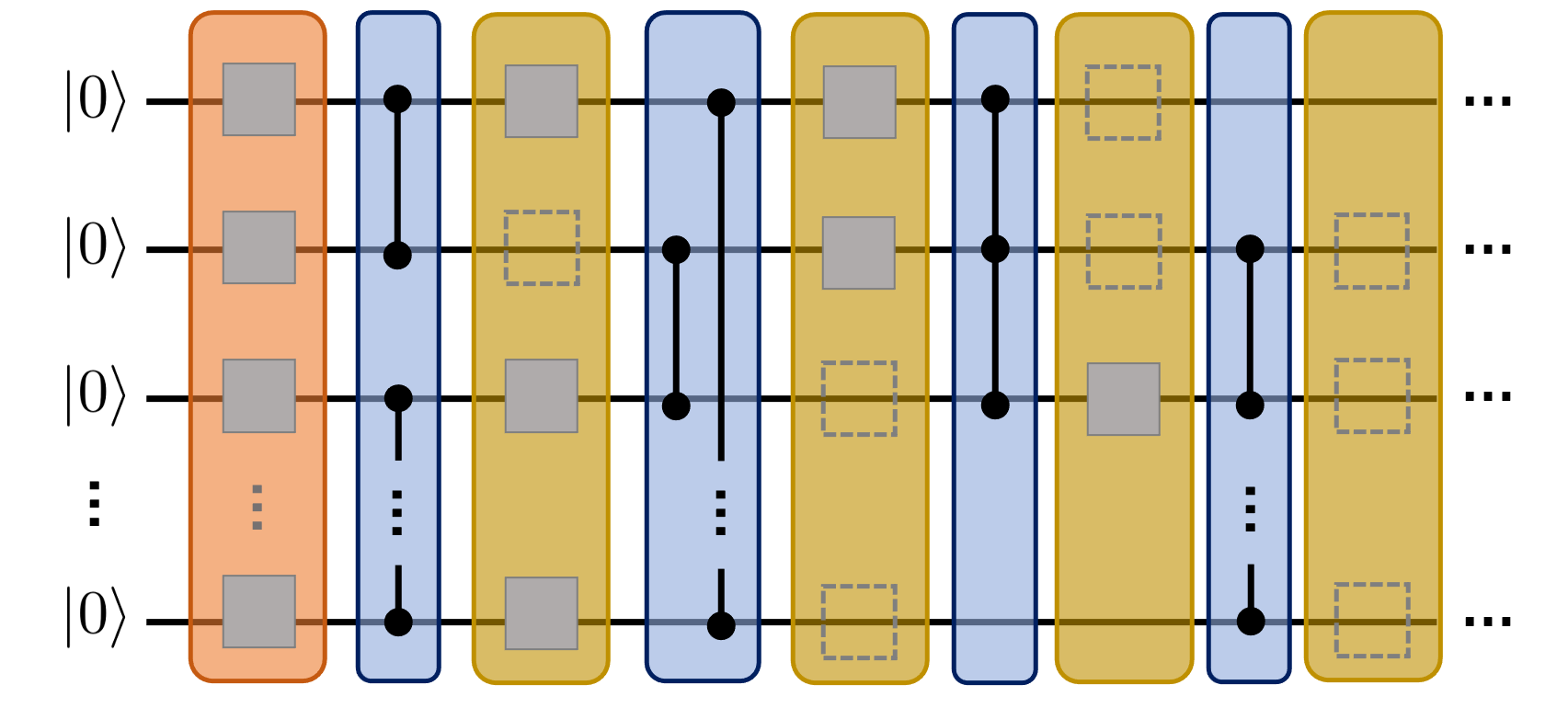}
\caption{\label{fig:generic_circuit} The format of every quantum circuit after proper optimization and transpilation. The first layer (in orange) has single-qubit gates on every qubit. Subsequent layers alternate between multi-qubit gates of varying sizes (in blue) or single-qubit gates on selected qubits (in yellow). Each layer can have at most one gate acting on each qubit, though not all qubits have to be acted on in every layer. For example, the dotted squares inside the yellow layers denote places where a single-qubit gate could be but isn't --- note how the third yellow layer does not allow for a gate on the last qubit because there was no multi-qubit gate on this qubit in the previous layer; if a gate was placed there, it would always be transferred to the previous layer or merged with the previous gate.
}
\end{figure}

While 1. and 2. are necessary conditions for the correct execution of the quantum circuit on the quantum device, 3. and 4. are the product of an optimization and transpilation routine that fully leverages the characteristics of the device. In particular, criterion 3. is achievable due to the capacity of the neutral-atom QPU to execute arbitrary single-qubit gates natively, while criterion 4. uses the fact that an MCZ where any of its qubits is in the $\ket{0}$ state is redundant, resulting in a circuit where the first gate acting on any qubit is always a single-qubit gate that does not commute with an MCZ. Finally, criterion 5. results from the simple observation that the state of a qubit experiencing only single-qubit gates can be trivially calculated classically and as such can be omitted from the circuit. A quantum circuit obeying all these criteria will always fit the outline of figure \ref{fig:generic_circuit}.

 When at least criteria 1. and 2. are met, the translation of a quantum circuit into a pulse sequence can be a naïve one --- each quantum gate has an equivalent pulse (or sequence of pulses) that can be executed in the same order as they appear in the circuit. However, knowing that there is a type of hardware channel dedicated to executing single-qubit gates and another for multi-qubit gates, one quickly realizes that time can be saved by operating the channels of different types in parallel. Already in a gate-level view, it is straightforward to realize that single-qubit gates and multi-qubit gates acting on mutually exclusive sets of qubits can be executed in parallel, a process we refer to as \textit{parallelization}. 
 
 However, we can go further if we leave behind the quantum-circuit model and switch to a pulse-level view of the sequence. The key observation here lies in the fact that multi-qubit gates are actually sequences of pulses targeting individual qubits either once or twice. Furthermore, a qubit is only involved in the process between the moments its first pulse starts and its second pulse ends --- in the case of the qubit with only one pulse (i.e. the $2\pi$-pulse), its involvement is constrained to the time the pulse is being applied. 
 
 Thus, if we consider the example of the CZ gate, we can see that the qubit on which the $2\pi$-pulse falls (let us call it $q1$) can be the object of a single-qubit gate while the other qubit ($q0$) is experiencing the $\pi$-pulses. This simultaneous execution of pulses is different from the concept of parallelization laid out above because, in this case, the qubits involved in the single-qubit gates are a subset of the qubits involved in the multi-qubit gate. Note how this process has no representation at the quantum-circuit level, where a gate is an indivisible block --- if we try to represent this in a quantum circuit, it will look as if the single-qubit gates on $q0$ are overlapping (or have been ``absorbed" by) the multi-qubit gate, which is why we name this process \textit{absorption}.
 
 \subsection{\label{sec:absorption}Absorption} 
 
\begin{figure*}
\includegraphics[width=\textwidth]{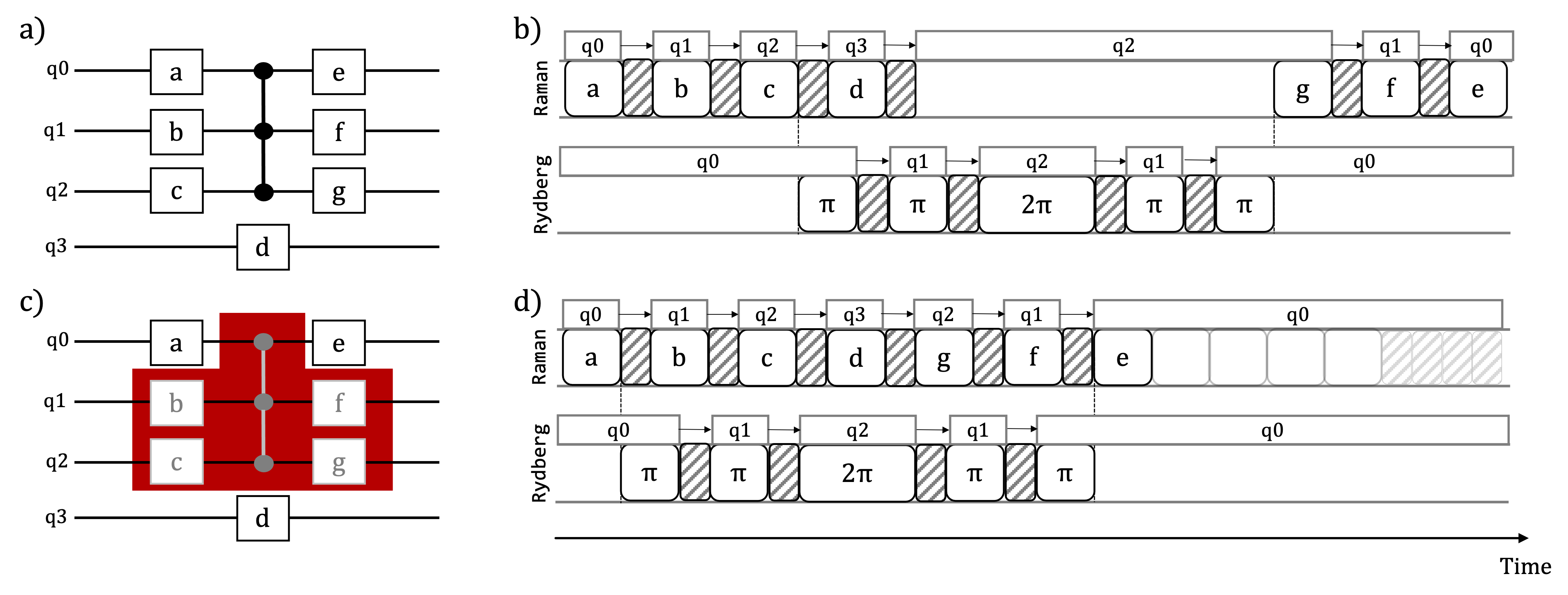}
\caption{\label{fig:absorption}An example on how \textit{absorption} can further reduce the time taken to execute a quantum circuit. a) The original circuit, consisting only of a CCZ gate with single-qubit gates preceding and succeeding all the qubits it acts on, plus a single-qubit gate on another qubit. b) The corresponding pulse sequence when only parallelization is allowed (i.e. gate \textit{d} is executed in parallel with the CCZ). Dashed areas correspond to retargeting times of the Raman and Rydberg channels c) The \textit{absorbed block}, showing which gates are chosen to be executed at the same time as the CCZ. d) The pulse sequence, now contemplating the absorption of the gates chosen in c). Compared to b), the time taken to execute four $\pi$-pulse blocks and four target blocks is saved.}
\end{figure*}

Consider the case depicted in figure \ref{fig:absorption}a, where a CCZ gate (an MCZ with three qubits) is preceded and succeeded by single-qubit gates on all the qubits it acts on, having also a single qubit gate which can be executed in parallel with any of the remaining gates. Knowing that the duration of a pulse is adjustable, we can assume for the sake of simplicity and without loss of generality that all pulses doing single-qubit gates have the same duration as a $\pi$-pulse, while the $2\pi$-pulse will last twice as long. Furthermore, let us assume that the time taken to change the target of a channel is the same on both channels. Under these assumptions, figure \ref{fig:absorption}b shows the result of converting the quantum circuit into the shortest possible pulse sequence, when only parallelization is allowed. In contrast, figure \ref{fig:absorption}d shows the same conversion but in the case where absorption is also allowed, which maximizes the amount of time both channels are simultaneously active and leads to a substantial reduction in the total duration of the sequence.

In figure \ref{fig:absorption}c, we show how absorption can be thought of as an agglomeration of single-qubit gates around $N-1$ qubits of an $N$-qubit MCZ gate, thus leaving out one qubit that we will call ``border" qubit --- the one that receives the first and last $\pi$-pulses and is thus the only one which cannot be acted on throughout the entire execution of the MCZ. Therefore, for an $N$-qubit MCZ gate, at most $2(N-1)$ single-qubit gates can be absorbed. 

Note that, although in this case the choice of the border qubit was irrelevant, it can always be made to maximize the number of absorbed single-qubit gates or to ensure some specific single-qubit gates are absorbed.

\subsection{Statement of the general problem}


The idea of absorption presented in section \ref{sec:absorption} shows how taking into account the execution of quantum gates at the pulse-level can lead to a substantial decrease in the time needed to execute a quantum circuit. Such a decrease in execution time is valuable because it minimizes the amount of decoherence incurred during the execution of a quantum circuit or, by the same token, permits a better use of the coherence time available on the device.

Therefore, our goal will be to create an algorithm that uses the idea of absorption to schedule the gates of a quantum circuit at the pulse-level with minimal execution time. This algorithm should handle any quantum circuit obeying the conditions laid out at the beginning of section \ref{sec:qc-to-ps} and produce a pulse sequence designed for a neutral-atom device with a Raman and a Rydberg channel.


We shall assume that any pulse on the Raman channel, as well as any $\pi$-pulse, can take at most $\delta_\pi$ time units to be executed, while a $2\pi$-pulse will take $2\delta_\pi$. Additionally, we shall assume that both the Rydberg and Raman channels take the same time $\delta_T$ to change their target.

\section{\label{sec:algo}The Algorithm}

First comes the absorption step, which consists in determining which single-qubit gates will be absorbed in the various MCZ gates in the circuit. For this step, we go through the multi-qubit gates of the circuit from left to right, making sure that a gate is only analyzed after all its predecessors. For each N-qubit MCZ, we choose up to N-1 of the qubits involved in that particular gate according to these priorities:
\begin{enumerate}
    \item The qubit has a single-qubit gate right before the current MCZ and is the border qubit of the MCZ previously analyzed for absorption.
    \item The qubit has single-qubit gates immediately before and after the MCZ.
    \item The qubit has a single-qubit gate before the MCZ.
    \item The qubit has a single-qubit gate after the MCZ.
\end{enumerate}

If a qubit does not meet any of these criteria, it means that it does not have any single-qubit gates to be absorbed and is thus not selected. On the other hand, if multiple qubits fulfill the same criterion, the choice between them can be arbitrary.

Then, we proceed to the ordering step, where we determine the order of execution of each absorbed block and unabsorbed single-qubit gate. Firstly, we identify all the unabsorbed single-qubit gates at the end of the circuit, which we shall refer to as \textit{final single-qubit gates}. Once all the predecessors of a final single-qubit gate have been executed, it is said to be \textit{ready for execution}.

We go through the absorbed blocks following the same order set during the absorption step. For each block, we follow these rules:
\begin{enumerate}
    \item The block is preceded by an unscheduled single-qubit gate:
    \begin{itemize}
        \item \textbf{Yes}. First schedule this single-qubit gate in parallel execution with the previous absorbed block, when possible, or on its own otherwise (which will be the case for the first block).
        \item \textbf{No}. Check if there is a final single-qubit gate ready for execution that is not succeeding the previous absorbed block; if so, schedule it for parallel execution with that block.
    \end{itemize}
    \item Schedule the absorbed block for execution.
    \item Check if any succeeding single-qubit gate is a final single-qubit gate and, if so, mark it as ready for execution.
\end{enumerate}

After these steps, there can be at most two unscheduled single-qubit gates: one not succeeding the last absorbed block and that can thus be scheduled in parallel with it, and another one succeeding it to be executed on its own. This implies that, in the worst case, the total duration of a circuit will be the duration of all the absorbed blocks plus two single-qubit gates. Furthermore, setting aside $\delta_\pi + \delta_T$ to realize each of the standalone single-qubit gates, we find that the duration of each absorbed block is always determined by the time taken to execute its MCZ, which means that the Rydberg channel is fully occupied executing one MCZ after the other without any pause. 

As such, this algorithm results in an optimal utilization of the Rydberg channel which forbids further reductions in the total sequence duration through a different scheduling of the pulses. Instead, all further improvements under these experimental conditions need necessarily to come from reductions in $\delta_\pi$ and $\delta_T$.

\begin{figure*}
\includegraphics[width=\textwidth]{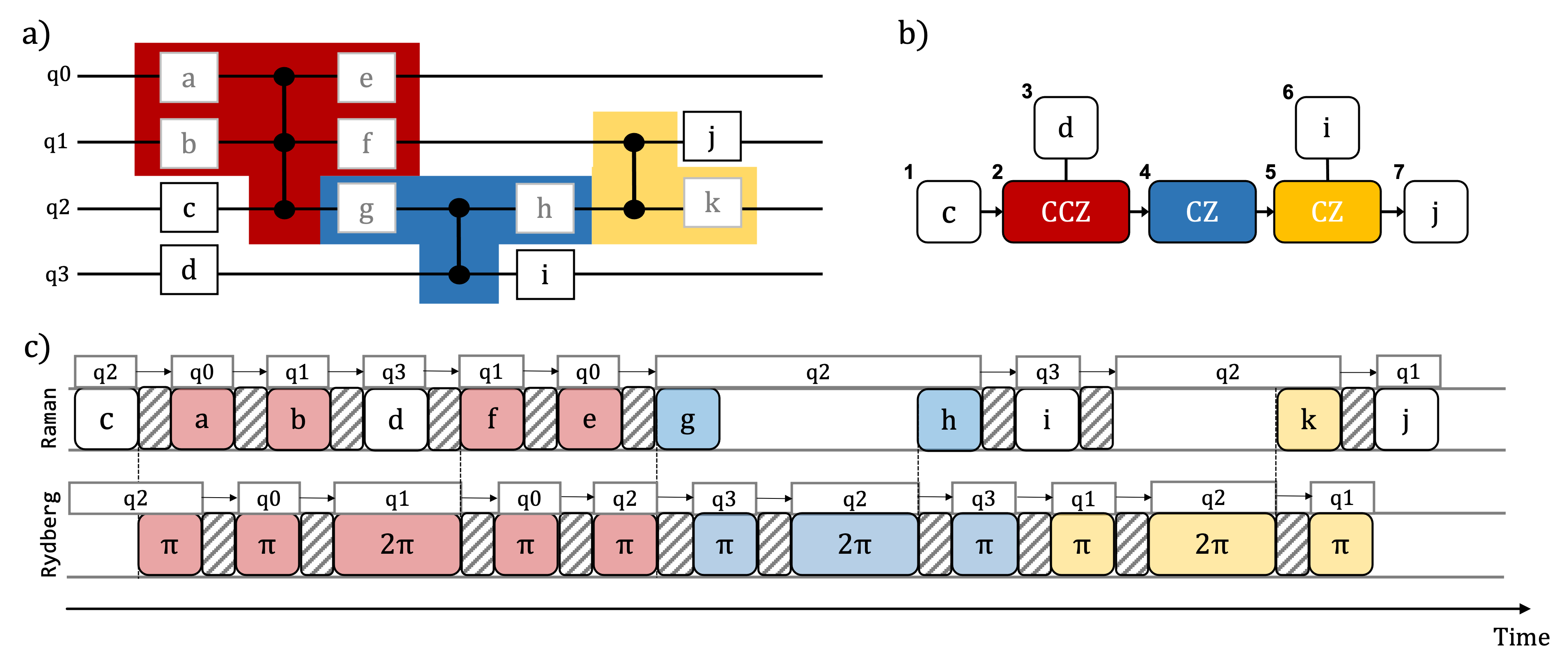}
\caption{\label{fig:algo_example}The two steps of the algorithm applied to a particular quantum circuit. a) The absorption step applied sequentially to each MCZ in the circuit, where each resulting absorbed block is represented by a different color. b) The ordering step, where the number next to each block indicates the order with each they were scheduled; the single-qubit gates connected to the top of absorbed blocks are the ones scheduled for parallel execution. c) The final pulse sequence, resulting from executing each block in b) from left to right. The pulses in the same absorbed block are marked with the color given to them in step a). Every instruction is executed as soon as possible while guaranteeing that a qubit is not acted on by two channels simultaneously (e.g. gate $i$ is scheduled for parallel execution with the yellow CZ but ends up starting even before the CZ because there are no absorbed single-qubit gates before it.)}
\end{figure*}

Figure \ref{fig:algo_example} illustrates the two steps of the algorithm being applied to a specific quantum circuit, along with the resulting pulse sequence.

\section{Benchmarks\label{sec:bench}}

The defining trait that sets the above-presented algorithm apart is the inclusion of pulse-level information in the process. Thus, to achieve a legitimate benchmark of our algorithm, we compare it against one that attempts to minimize the total sequence duration with the information available at the quantum-circuit level.

\subsection{The gate-level scheduling algorithm}

Since gate parallelization is still possible at the gate level, we designed a custom scheduling algorithm that attempts to minimize the total sequence duration by maximizing the number of single-qubit gates executed in parallel with the MCZ gates.

Starting from a quantum circuit of practical relevance, we first observe that the problem can be simplified to the scheduling of its MCZ gates, as long as the execution of each MCZ gate is always preceded by the execution of the single-qubit gates immediately before it. We combine this fact with the observation that, if we organize the circuit in layers following the scheme of figure \ref{fig:generic_circuit}, we can schedule the MCZ gates within the same layer in any order without risk of a conflict (i.e. one gate depending on the execution of another). Furthermore, doing so maximizes the chances that the single-qubit gates preceding an MCZ gate are scheduled in parallel with previously scheduled MCZ gates.

As such, our gate-level scheduling algorithm boils down to going through each multi-qubit gate layer in the circuit and, for each layer, scheduling all its multi-qubit gates before proceeding to the next layer. Within the same layer, we identify the last time each gate's qubits were involved in a scheduled gate, take the maximum value among the qubits of each gate and use that value to sort the MCZ gates in ascending order. In the event of a tie, we schedule the gates with the least preceding single-qubit gates first (in an arbitrary order, if there is still more than one).

After all multi-qubit gates are scheduled, we are left with the final single-qubit gates of the circuit. Knowing that they can be parallelized with any MCZ after their last MCZ gate was scheduled, we look for the earliest available slot to schedule them from that point on, starting with the gates that can be executed the earliest.

\subsection{Data generation and analysis}

We generate a series of random quantum circuits with variable numbers of qubits and multi-qubit layers (which we will refer to as layers from here on). We start out with an empty circuit with a set number of qubits and apply a non-diagonal single-qubit gate to every qubit. Then, we randomly pick, with equal probability, whether to apply a CZ or a CCZ gate; depending on the pick, we randomly choose two or three qubits to apply the chosen gate, followed by non-diagonal single-qubit gates. We repeat this process at least until all qubits in the circuit are involved in at least one MCZ gate and stop once a set minimum number of MCZ gates is reached. We repeat this process for various numbers of qubits and MCZ gates to create a sufficiently large and diverse data set.


We then pass the resulting quantum circuits through a transpilation routine that first maps the qubits to vertices in a triangular lattice layout with next to nearest-neighbour connectivity, meaning that a qubit (away from the edges) is connected to the 12 qubits closest to it. Then, it performs qubit routing to enable operations between physically disconnected qubits, which is achieved through the addition of SWAP gates. Since this process changes the composition of the quantum circuit, we finish by decomposing the non-native gates, optimizing to reach a quantum circuit of practical relevance and counting the final number of layers.

All the circuits are then grouped according to their qubit and layers numbers, and scheduled by the two scheduling algorithms under the same conditions. In particular, we fix $\delta_{\pi} = \delta_{T}$, which allows us to express the time gained by the pulse-level scheduling approach in units of $\delta_\pi$, as is done in figure \ref{fig:benchmarks}.

\subsection{Results}

\begin{figure}
\includegraphics[width=\columnwidth]{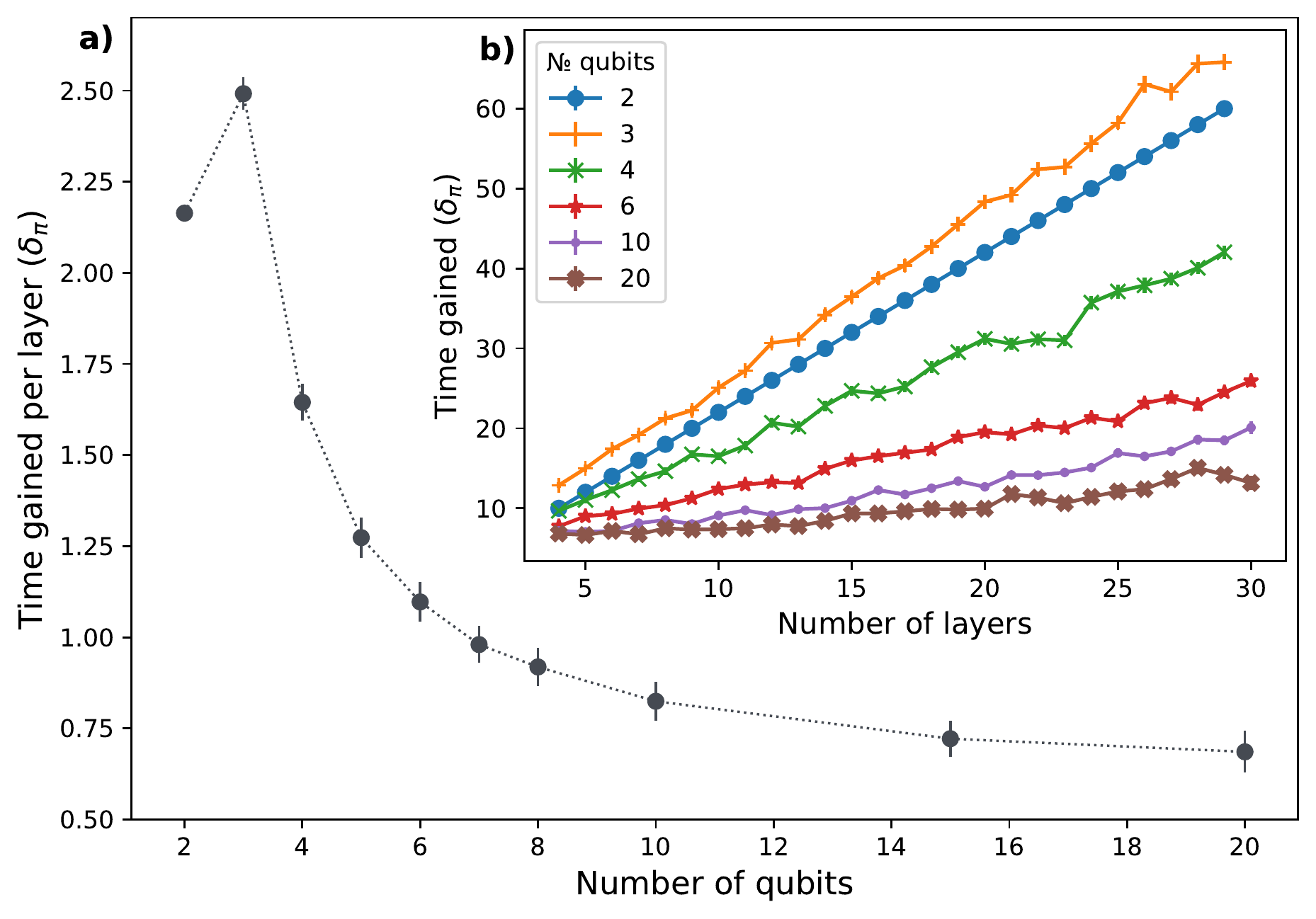}
\caption{\label{fig:benchmarks} The time gained (in units of the $\pi$-pulse duration, $\delta_\pi$) by the pulse-level scheduling algorithm when compared to an algorithm considering only gate-level information, for a series of random quantum circuits of different sizes in number of qubits and layers. Each data point in panel b) represents the average total time gained over multiple random circuits (generally around 75) with the same number of qubits and layers. For a given number of qubits, each point is then divided by its number of layers and averaged again to yield the average time gained per layer expressed in panel a).
}
\end{figure}

The benchmark results compiled in figure \ref{fig:benchmarks} exhibit how the time gained from using the pulse-level scheduling algorithm is directly proportional to the depth of the circuit --- more specifically, to the number of multi-qubit layers, as illustrated in panel b). Furthermore, the proportionality constant between time gained and the number of layers decreases monotonically with the number of qubits, from circuits with three qubits onward, as displayed in panel a). These results make intuitive sense: the larger the number of qubits in a circuit is, the more likely one is to find multiple MCZ gates in the same layer, which allows for parallelization of single-qubit gates with MCZ gates and thus reduces the relative overhead incurred by the gate-level scheduler.

In fact, having more than one multi-qubit gate per layer only becomes possible from four-qubit circuits on, which explains the sharp decline in time gained when compared to the three qubits case. The slight increase in time gained between two- and three-qubit circuits is also noteworthy. It is explained by the fact that CCZ gates only start being added when there are three or more qubits and, when its preceding single-qubit gates cannot be parallelized, each CCZ ends up idling the Rydberg channel longer than a CZ. By the same token, the two-qubit circuits all end up having a single CZ per layer, surrounded by single-qubit gates, which means that for scheduling purposes all two-qubit circuits with the same number of layers are identical --- as a consequence, we get all the points lying perfectly along the two-qubit line of panel b).

As for the aspects that were fixed during the benchmarks, it is worth speaking of the effect of the connectivity and the choice of multi-qubit gates. The connectivity influences the number of gates in a circuit that can be executed without routing. The smaller it is, the more routing will need to take place to enable the execution of a given circuit. We expect that the addition of SWAP gates leads to the spreading of the multi-qubit gates over more layers, decreasing the number of gates per layer and thus increasing the average time gained per layer. The choice of multi-qubit gates plays a similar role: here, we went with only with CZ and CCZ gates, but the larger the gates are, the higher demands on routing will be --- furthermore, this effect will be enhanced by the need to decompose the gates when the connectivity does not permit their native execution.

\section{Conclusion}

We have showed how the pulse-level implementation of the multi-qubit gates in neutral atom device architectures allows for the simultaneous execution of single- and multi-qubit gates acting on overlapping sets of qubits, in a mechanism we dubbed absorption. 

Using the concept of absorption as a foundation, we designed an algorithm to schedule the execution of a quantum circuit as a pulse sequence on a neutral-atom device with a single Raman and Rydberg channel. For any quantum circuit of practical relevance, we observe that the total duration of a pulse sequence is always bounded by the number of pulses in the Rydberg channel and that the developed algorithm schedules all the Rydberg pulses in an uninterrupted stream. As such, we conclude that the algorithm results in an optimal utilization of the available resources that cannot be surpassed by a different scheduling strategy.

Finally, we benchmarked our algorithm against a custom scheduler that attempts to maximize the parallelization of single- and multi-qubit gates. We found that the time gained by the pulse-level scheduler generally decreases as the number of qubits in the circuit increases, since circuits with more qubits offer more chances for parallelization. Furthermore, we found that the time gained is proportional to the depth of the circuit, making the total time saved by the pulse-level scheduling algorithm more pronounced for longer quantum circuits.

\begin{acknowledgments}
We thank Laurent Ajdnik and Louis-Paul Henry for useful discussions.
This project has received funding from the European Union’s Horizon 2020 research and innovation program under grant agreement No 968614.
\end{acknowledgments}

\bibliography{biblio}

\end{document}